\documentstyle[aps,prb,preprint,epsf,epsfig,amsmath]{revtex}
\newcommand{\beq}{\begin{equation}}
\newcommand{\eeq}{\end{equation}}

\newcommand{\vecg}[1]{\mbox{${\boldsymbol #1}$}}
\newcommand{\vech}[1]{\mbox{\scriptsize${\boldsymbol #1}$}}
\newcommand{\veca}[1]{\mbox{\bf #1}}
\newcommand{\vecb}[1]{\mbox{\bf\scriptsize #1}}
\newcommand{\inst}[1]{\mbox{$^{#1}$}}

\begin{document}
\draft
\title{Dipolar Interactions in Superconductor-Ferromagnet
Heterostructures}
\author{Jaime E. Santos\inst{1} \and Erwin Frey\inst{1,2} \and 
Franz Schwabl\inst{1}}
\address{1 - Institut f\"ur Theoretische Physik T34, 
Physik-Department der 
TU M\"unchen,
James-Franck-Stra\ss e, 85747 Garching, Germany\\
2 - Lyman Laboratory of Physics,
Harvard University, Cambridge, MA 02138, USA\\
email: {\rm jesantos@physik.tu-muenchen.de} \\
email: {\rm frey@cmts.harvard.edu}\\
email: {\rm schwabl@physik.tu-muenchen.de}
} 
\maketitle
\date{}
\newpage
\begin{abstract}
We consider a simple model for a superlattice
composed of a thin magnetic film placed between
two bulk superconductors. The magnetic film is modelled
by a planar but otherwise arbitrary distribution of magnetic
dipoles and the superconductors are treated in the
London approximation. Due to the linearity of the problem, we
are able to compute the magnetic energy of the film in the presence
of the superconductors. We show that for wavevectors which are much
larger than the inverse London penetration depth, 
the magnetic energy is
unchanged with respect to the film in free space, whereas in the
case of small wavenumbers compared to the inverse
London penetration depth,
the magnetic energy resembles the energy of a
distribution of magnetisation in a two dimensional space.
Possible experimental applications of these results are discussed.
\end{abstract}
\pacs{75.70.-i, 74.80.-g} 
\section{Introduction}
\label{secA}
The interplay between superconductivity and ferromagnetism 
in bulk materials has been the subject of 
active research since 1957 when Ginzburg 
\cite{Ginz57} published a paper in which he considered the 
effect of the field created by a bulk distribution 
of magnetisation on a superconductor, which was described 
by the London equations \cite{London61}. 
He has concluded that for a ferromagnetic induction field of the  
sample larger than its superconducting critical field,
this field would destroy superconductivity, 
but he also pointed out that in thin films or wires where the
induction field is much smaller (due to demagnetisation effects)
and the critical field higher (due to the small diamagnetic energy)
than in bulk superconductors, it should be possible to observe 
the coexistence of the two phenomena. Experiments 
carried out by Mathias et al. \cite{Mathias58} on Lanthanum 
with several rare-earth paramagnetic impurities dissolved at 
low concentrations suggested that the interaction 
responsible for the depletion
of the superconducting critical temperature of Lanthanum 
is the exchange interaction between the paramagnetic
impurity spins and the superconducting electrons.  This interaction
induces an effective ferromagnetic 
interaction between the (antiparallel)
spins in the Cooper pair, which tends to destroy 
it and hence destroy 
superconductivity. Anderson and Suhl \cite{Anderson59} have shown 
that the RKKY interaction between the 
ferromagnetic spins due to the conduction electrons  
is significantly reduced in the superconducting state,
but pointed out that ferromagnetism could coexist
with superconductivity if the ferromagnetic atoms formed small
domains. The dependence of the superconducting critical temperature
on the concentration of magnetic impurities due to exchange
scattering of electrons from these impurities
was addressed with the microscopic theory of 
superconductivity by Abrikosov and Gor'kov \cite{Abrikosov61}. 
de Gennes and Sarma \cite{Genn63} 
have estimated that tipically,
the exchange interaction between localised moments
and superconducting electrons would be $10^2-10^3$ larger 
than the dipolar interaction considered by Ginzburg.
The detailed form of the Landau-Ginzburg theory of 
ferromagnetic superconductors was worked out by Suhl \cite{Suhl78}.
Despite their conflicting character, 
superconductivity and ferromagnetism
are seen to coexist in bulk systems, e.g.  
in HoMo$_6$S$_8$ and ErRh$_4$B$_4$ \cite{Adv85}. 

Another system in which the coexistence of superconductivity and 
ferromagnetism has been observed in the bulk is the nuclear 
magnet AuIn$_2$ \cite{Herr95a,Herr95b,Rehmann97,Kulic97}. 
This compound shows a superconducting phase transition 
at $T_c=207 mK$ and an ordering transition 
to a ferromagnetic state 
at an even lower temperature $T_M=35 \mu K$. 
This particularly low temperature can be explained by the
weakness of the interaction between the nuclear spins
(which is primarily due to an indirect exchange via the conduction
electrons). 

On a different perspective, the development of epitaxial 
growth of crystals has permitted the 
creation of artificial superlattices 
composed of superconducting and ferromagnetic materials, e.g.  
Fe/V, Ni/V, Ni/Mo, EuS/Pb, EuO/Al and Nb/Gd
\cite{Jin89,Zinn83,Wong86,Tedrow86,Takahashi87,Millis88,Strunk94}. 
In these superlattices, one can experimentally 
study the interaction 
between superconductivity and ferromagnetism 
when these two effects 
occur in neighbouring spatial regions and also study the supression 
of superconductivity as a function of the relative proportion 
(i.e. layer tickness) of the two materials. More recently, the 
cuprates  RuSr$_2$GdCu$_2$O$_{8-\delta}$ 
\cite{Bauern95A,Bernhard99,Pringle99,Hadjiev99} and 
RuSr$_2$Gd$_{1+x}$\-Ce$_{1-x}\-$Cu$_2$O$_{10}$ \cite{Bauern95B}
have been found to show superconductivity and a ferromagnetism 
below their critical temperatures, $T_c=15$-$40K$ for 
RuSr$_2$Gd\-Cu$_2$O$_{8-\delta}$ and $T_c=37K$ (for an optimal 
$x=0.2$) for  RuSr$_2$Gd$_{1+x}$\-Ce$_{1-x}\-$Cu$_2$O$_{10}$,
the Curie temperatures for magnetic 
ordering being $T_M\approx 133K$
for the first compound and $T_M\approx 100K$ for the second. 
The experimental analysis shows that these materials, 
like all cuprates, 
have a layered structure and that 
superconductivity and ferromagnetism 
seem to occur in different layers. 
However, a detailed analysis 
has been hindered by difficulties with 
the growth of single crystals \cite{PBraun}.

\begin{figure}[htbp]
\centerline{\epsfxsize 0.75\columnwidth \epsfbox{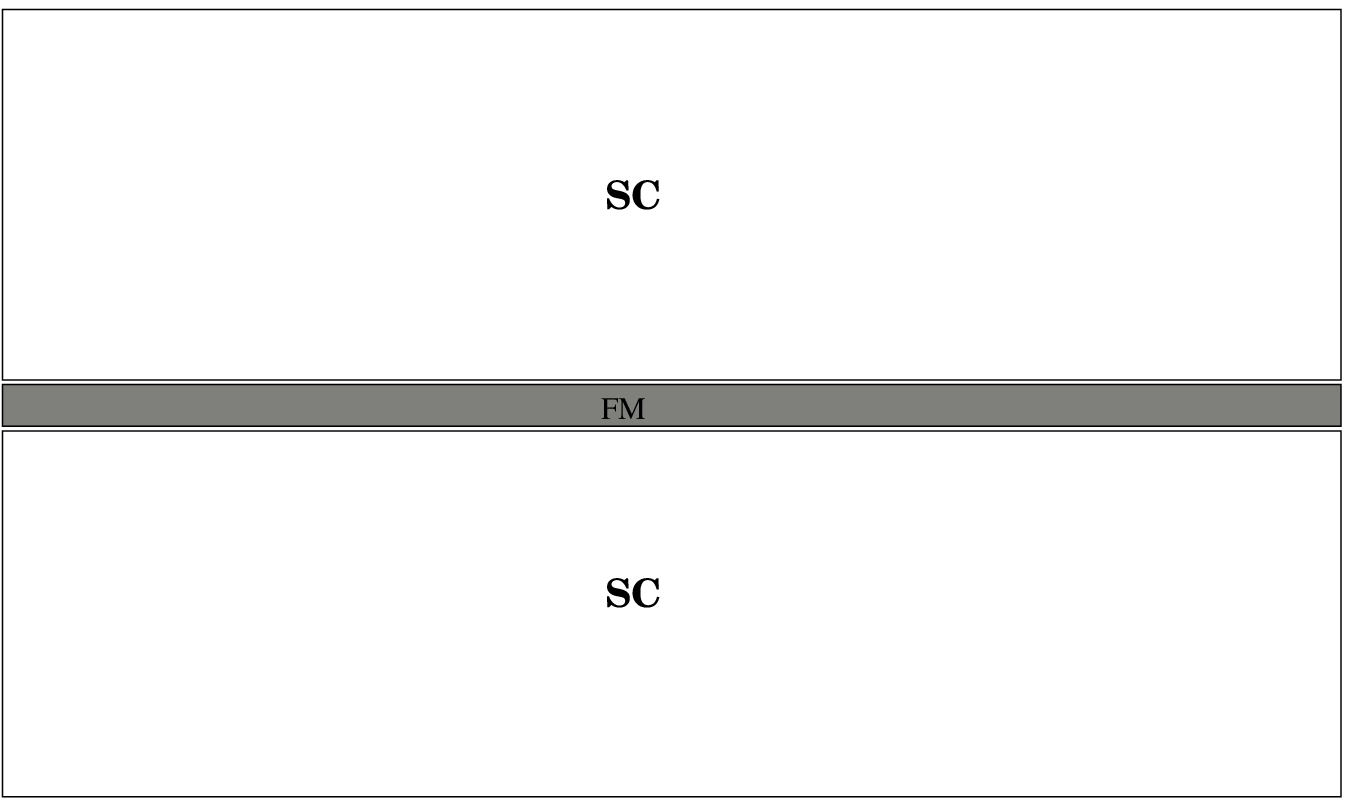}}
\vspace{3mm}
\caption{\label{F1}}
\end{figure}

Motivated by such experiments, 
in which magnetism and 
superconductivity are seen to occur in different 
spatial regions of the studied materials, we wish 
to address the problem of a thin ferromagnetic layer, 
placed between two bulk superconducting layers (see fig. \ref{F1}),
in which the thickness of the superconducting layers is much larger
than the London penetration depth of the superconducting material
and the thickness of the ferromagnetic layer is very small compared
to this quantity, which is a condition that can be easily obtained
with the modern techniques of epitaxial growth \cite{Jin89}.
In this limit, the results obtained can also be applied to 
superlattices of the two materials, given that the ferromagnetic
layers are decoupled from one another.

We consider in this paper a simple model system 
composed of a very thin ferromagnetic film, 
with an arbitrary distribution 
of magnetisation in the plane of the film,  which is placed 
in a spatial gap of size $2d$ between two semi-infinite 
superconductors described by the London equations. The film is 
coupled to the superconductors by the 
electromagnetic interaction, i.e. we
neglect the proximity effect \cite{Genn69}
and we consider the Josephson current \cite{Josephson64}
flowing between the two superconductors 
to be zero (the limitations of these
approximations will be discussed in section \ref{secD}).
Having made the approximations indicated above, we 
are able to solve the problem exactly, 
by firstly considering the simpler problem 
of a single dipole in the spatial gap
and then superimposing the different solutions, 
due to the linearity of the London equations. One can then 
compute the dipolar energy of the
distribution of magnetisation. It turns 
out that for wave-vectors much larger 
than the inverse London penetration 
depth the form of the dipolar energy
in momentum space is unchanged by the 
presence of the superconductors.
On the other hand, for wave vectors much 
smaller than the inverse London
penetration depth, the dipolar energy 
in momentum space resembles the energy
of a distribution of dipoles in a two 
dimensional space. This behaviour can
be traced to the Meissner effect 
which confines the magnetic flux lines within
the spatial gap. 

One can think of several possible ways to 
detect the effects of this change of behaviour 
of the dipolar interaction at low wave vectors.
If one were able to choose the materials composing
a layered geometry of 
superconductor/fer\-ro\-magne\-tic film/superconductor 
in such a way
that the Curie transition temperature of the magnetic film
to the ferromagnetic state is
lower than the critical temperature 
$T_c$ of the superconductor, one should be able to
measure the critical properties of the system at the 
paramagnetic-ferromagnetic transition, in particular 
such quantities as the specific heat and the 
magnetic susceptibility, with the
superconductors already displaying the Meissner effect
and therefore with a modified form of the dipolar energy.

It was shown by Aharony and Fisher \cite{Aharony73} 
that in a $d$-dimensional system,
a $d$-dimensional dipolar interaction (such as the one occuring
in a bulk tridimensional ferromagnet or in the layered geometry
superconductor/ferromagnet/superconductor) 
is a relevant interaction (in the sense of RG) 
near a paramagnetic-ferromagnetic
transition, leading to a crossover between
the critical exponents of the short-range ferromagnet
and the critical exponents of the dipolar system
when one approaches the critical temperature
\cite{Bruce74,Nattermann76,Bruce76,MASantos80,Kogon82,FreyS}. 
Pelcovits and Halperin \cite{Pelcovits79}
have also shown that in the case of a $d$-dimensional
systems with a $d+1$-dimensional dipolar interaction (such as
the one ocurring in a thin magnetic film film in free space) 
the universality class of the dipolar system is the same as above.
This is due to the fact that, at the fixed point,
the `effective' (renormalised) dipolar coupling constant
is infinite, making the susceptibility of the system
independent of the longitudinal degree of freedom
of the magnetisation, which is the one sensitive to the nature 
of the dipolar interaction. However, in real
systems, measurements are not taken exactly at
the critical point and one always probes the crossover
region. In this region, the dipolar coupling constant is finite
and one should be able to detect the distinct character of the
transition if the ferromagnetic film is included in a 
layered geometry with superconductors or if the film 
is grown in a non-superconducting substrate, 
due to the different character of the dipolar interaction 
at small wavevectors. 
The ideal experiment to detect such a distinction
would presumably be a measurement 
of the longitudinal susceptibility using polarised
neutrons \cite{Lovesey84,Koetzler86,Boeni91,Maier98,note1}.
Experiments done with films of EuS/SrS
grown on a Si substrate have shown \cite{Zinn83} that the 
low Curie temperature of EuS (16.5 K) is further reduced 
in these geometries. The authors of Ref. \cite{Zinn83}
have also performed experiments
with films of EuS/Pb, probing the transition between 
the superconducting state and the normal state 
in the Pb layer as a function of the applied
magnetic field. Therefore, EuS stands as a good candidate
for a  material to be used in the ferromagnetic layer.
It has the further advantage of being an insulator (see below).

Another possibility would be the study of spin-spin 
correlation functions in a magnetic film in the 
ordered phase and outside the critical region. 
Indeed, Kashuba \cite{Kashuba94} has shown that the static 
correlation functions of an XY model with 2d dipolar interactions 
would display a behaviour analogous to that of the dynamic correlations 
functions of the stochastic process described the KPZ 
e\-qua\-tion in 1+1 dimensions \cite{KPZ86}, for which the form
of these correlation functions is known. An adequate experiment 
to probe these correlation functions at low momentum compared with
the inverse London penetration depth  (where such length
is typically of the order of a thousand
Angstroms) would presumably be low-angle neutron 
scattering from the magnetic  fluctuations in a layered geometry. Other 
possible experiments which could probe
the magnetic properties of the system in the ordered phase
would be the use of the 
magnetooptical Kerr effect or of the Faraday effect 
on samples with a single magnetic layer to image such a layer. 

The structure of this paper is as follows:
in section \ref{secB}, we define our model in terms of the
geometry of the system and the equations which describe
it. We also describe the type of boundary conditions we
have to consider. In section \ref{secC}, we present the solution of
the equations for a single dipole and construct the solution
for a general in-plane distribution of magnetisation by linear
superposition. In section \ref{secD}, we compute the dipolar energy
of the distribution of magnetisation 
and discuss the physical limitations 
of the model we have considered. 
Finally, in section \ref{secE}, we present our conclusions.

\section{Geometry of the model and relevant equations}
\label{secB}
\begin{figure}[htbp]
\centerline{\epsfxsize 0.75\columnwidth \epsfbox{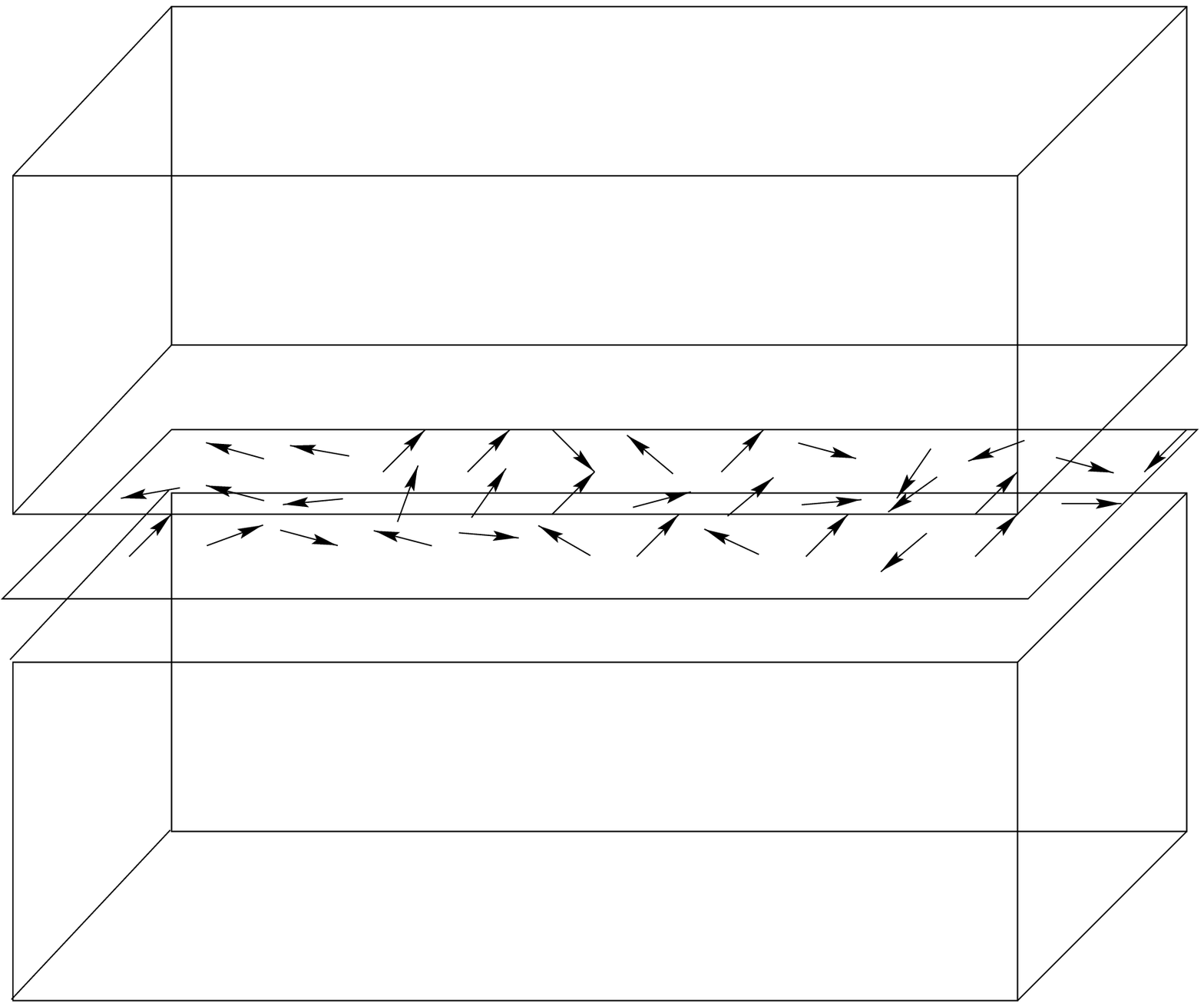}}
\vspace{3mm}
\caption{\label{F2}}
\end{figure}

The geometry of the model is as follows: an infinite
distribution of in-plane magnetisation 
is placed in the plane $z=0$. 
This distribution is constituted by single magnetic dipoles, 
placed in an arbitrary fashion with respect 
to one another (see fig. \ref{F2}). The in-plane
constraint implies that all the dipoles point 
in a direction within
the plane. Above and below the distribution are 
two bulk superconductors
which extend from $z=d$ (respectively $z=-d$) 
to $z=\infty$ (respectively
$z=-\infty$). The spatial gap with size 
$2d$ is supposed to be filled with
an insulator with magnetic permitivity $\mu_0$. The
two superconductors are identical 
and have a magnetic permitivity $\mu$
(i.e. they are paramagnetic, with relative permitivity
$\mu_r=\mu/\mu_0$). These supercondutors are
described by the London equations (see below), 
which imply a linear relation
between the current and the magnetic field.

This linear relation allows us to 
consider instead a simpler problem,
the one of a single magnetic dipole, 
placed at the origin of the coordinate
system and oriented along the $x$ axis. 
Once this problem has been solved,
one can construct the solution 
for the general case simply by using translational
and rotational invariance in the plane 
and by adding the different solutions.
The linearity of the equations will 
guaranty that the linear combination is
also a solution. Furthermore, a uniqueness theorem 
proved by London \cite{London61}
guaranties that this solution is unique.

In the spatial gap, the system is 
described by the following equations
\cite{Jackson99},
\begin{eqnarray}
\label{eq1}
\nabla\times\veca{h}&=&\veca{0}\\
\label{eq2}
\nabla\cdot\veca{b}&=&0\\
\label{eq3}
\veca{b}&=&\mu_{0}(\veca{h}+\veca{m})\, ,
\end{eqnarray} 
where the first two equations 
are the Maxwell equations for the magnetic
field $\veca{h}$ and the magnetic induction $\veca{b}$ 
and the third equation
is the constitutive relation between the two. 
For the case of a single
dipole oriented along the $x$ axis, 
the magnetisation $\veca{m}(\veca{r})=
m\,\hat{\vecg{x}}\,\delta^3(\veca{r})$, 
where $m$ is the magnitude of 
the magnetic dipole. 

The su\-per\-con\-duc\-tors are  
described by the equations
\begin{eqnarray}
\nabla\times\veca{h}&=&\vecg{\jmath}
\label{eq4}\\
\veca{b}&=&-\Lambda\nabla\times \vecg{\jmath}^s
\label{eq5}\\
\veca{b}&=&\mu\veca{h}
\label{eq6}
\end{eqnarray}
where the first equation is the Maxwell 
equation which relates the magnetic
field with the `free' current, 
the second relation is the second 
London equation \cite{London61} 
which relates the supercurrent $\vecg{\jmath}^s$ 
with the magnetic induction and the third 
equation is the constitutive relation
between the magnetic induction and 
the magnetic field. The constant 
$\Lambda$ is dependent on the 
type of the superconductor. In a static situation
such as the one we are considering, 
the electric field $\veca{e}=\veca{0}$ in the
superconductor and the total current 
$\vecg{\jmath}=\vecg{\jmath}^s$, i.e. there is 
no normal component of the current.
 
Subs\-ti\-tu\-ting equa\-tions 
(\ref{eq5}), (\ref{eq6}) in equa\-tion (\ref{eq4}),
one obtains
\begin{eqnarray}
\label{eq7}
\nabla^2\vecg{\jmath}-\lambda_L^{-2}\vecg{\jmath}&=&\veca{0}
\\
\nabla^2\veca{b}-\lambda_L^{-2}\veca{b}&=&\veca{0}
\label{eq8}
\end{eqnarray}
where $\lambda_L=(\Lambda/\mu)^{1/2}$ 
is the London penetration depth and 
where equation (\ref{eq8}) 
follows from taking the curl of (\ref{eq7}) 
and using (\ref{eq5}) and where we have 
used the fact  that $\nabla\cdot 
\vecg{\jmath}=0$ (equation of continuity) and 
$\nabla\cdot \veca{b}=0$.
These two equations show that 
the magnetic flux density and the supercurrent
penetrate a layer of thickness $\lambda_L$ at the surface of the 
superconductor (Meissner effect). 

These equations have to be suplemented by 
boun\-da\-ry conditions at the surface of the superconductors. 
These conditions are the continuity 
of the normal component of $\veca{b}$, of the tangential components 
of $\veca{h}$ and of the normal component of the current
$\vecg{\jmath}$ at the boun\-da\-ry surfaces 
of the two superconductors 
\cite{London61}.  
If one chooses $\vecg{\jmath}=
\nabla\times(g\hat{\vecg{z}})$, where $g(\veca{r})$ 
is a solution of 
\begin{equation}
\nabla^2 g-\lambda_L^{-2}g=0
\label{eq9}
\end{equation}
then one can satisfy the equations (\ref{eq7}) and (\ref{eq8})
and the boun\-da\-ry condition $\jmath_z=0$ 
at $z=\pm d$. Notice that this
choice implies that 
$\jmath_z=0$ throughout the material which is physically
reasonable, since $\jmath_z=0$ at the surfaces $z=\pm d$ and also 
for $z=\pm \infty$. 

On the other hand, in the spatial gap, 
we obtain from equation (\ref{eq1})
$\veca{h}=-\nabla\Phi_M$. 
Substituting this result in equation (\ref{eq3})
and using equation (\ref{eq2}), we obtain
\begin{equation}
\nabla^2\Phi_{M}=\nabla\cdot(m\,\hat{\vecg{x}}\,\delta^3(\veca{r}))
\label{eq10}
\end{equation}
which is Poisson's equation. Since we know the solution of
this equation in free space (i.e. in the absence of the 
superconductors) to obtain the solution in this case, we can write
\begin{equation}
\Phi_{M}(\veca{r})=\frac{m \rho \cos\phi}{4\pi(\rho^2+z^2)^{3/2}}+
\chi(\rho,\phi,z)
\label{eq11}
\end{equation}
where the first term on the rhs is the solution in free space
and the function $\chi(\rho,\phi,z)$ 
is a solution of the 
Laplace equation, $\nabla^2\chi(\rho,\phi,z)=0\,$ 
and where we have used cylindrical polar
coordinates for later convenience.
 
Therefore we need to solve 
the modified Helmholtz equation (\ref{eq9})
in the superconductors and the 
Laplace equation for $\chi$ in the gap
and then fit the two solutions 
using the continuity conditions for $\veca{b}$
and $\veca{h}$ at the boundary.

We can further simplify the problem if we notice that the
system is invariant under a $\pi$  rotation around the $x$ axis.
This invariance imposes the conditions
\begin{eqnarray}
\label{eq12}
g_-(\rho,\phi,z)=-g_+(\rho,-\phi,-z)\\
\label{eq13}
\chi(\rho,\phi,z)=\chi(\rho,-\phi,-z)
\end{eqnarray}
where $g_+$ (respectively $g_-$) 
is the solution of the Helmholtz equation in
the upper (respectively lower) superconductor.
Since the magnetic flux in the superconductor is given in terms
of $g$ by
\begin{eqnarray}
\label{eq14}
b_{z}&=&\mu\lambda_L^2\left(\frac{1}{\rho}\frac{\partial}
{\partial\rho}\left(\rho\frac{\partial g}{\partial \rho}\right)+
\frac{1}{\rho^2}\frac{\partial^2g}{\partial\phi^2}\right)\\
\label{eq15}
b_{\rho}&=&-\mu\lambda_L^2\,
\frac{\partial^2g}{\partial\rho\partial z}\\
\label{eq16}
b_{\phi}&=&-\frac{\mu\lambda_L^2}{\rho}
\,\frac{\partial^2g}{\partial\phi\partial z}
\end{eqnarray} 
and we have, in the spatial gap 
$\veca{b}=\mu_0\,(-\nabla\Phi_M(\veca{r})+
m\,\hat{\vecg{x}}\,\delta^3(\veca{r}))$, 
then the continuity conditions for 
$\veca{b}$ and $\veca{h}$ imply that at $z=d$,
\begin{eqnarray}
\label{eq17}
\left.\frac{\partial\Phi_M}{\partial z}\,\right|_{z=d}
&=&\mu_r\lambda_L^2\left(
\frac{\partial^2g_+}{\partial^2 z}-
\lambda_L^{-2}g_+\right)_{z=d}\\
\label{eq18}
\left.\frac{\partial\Phi_M}{\partial \rho}\,\right|_{z=d}
&=&\lambda_L^2\left(
\frac{\partial^2g_+}{\partial\rho\partial z}\right)_{z=d}\\
\label{eq19}
\left.\frac{\partial\Phi_M}{\partial \phi}\,\right|_{z=d}
&=&\lambda_L^2\left(
\frac{\partial^2g_+}{\partial\phi\partial z}\right)_{z=d}
\end{eqnarray}
where we have used the fact that $g_+$ is a solution
of the Helmholtz equation and that, in the superconductor, 
$\veca{b}=\mu\veca{h}$. 
A similar set of conditions is valid at $z=-d$ 
but they are trivially related to these conditions by
equations (\ref{eq12}) and (\ref{eq13}).

The above equations and boundary conditions 
are sufficient to determine
the solution of the problem within the London approximation. 
  
\section{The single dipole solution and the general solution for
an arbitrary distribution of magnetisation}
\label{secC}
We concluded in the previous section that
in order to find the field and current distributions for the
case of a single dipole, one needs to find a joint solution of 
the Laplace and Helmholtz equations, which satisfies the 
appropriate boundary conditions (\ref{eq17})-(\ref{eq19}).
Such a solution can be most easily found 
using cylindrical polar coordinates
and is given in terms of Fourier-Bessel integral transforms by
\begin{eqnarray}
\label{eq20}
\Phi_M(\rho,\phi,z)&=&\frac{m}{4\pi}
\left(\frac{\rho\cos\phi}{(\rho^2+z^2)^{3/2}}
+\int^{\infty}_0\,dk\,k\,J_1(k\rho)
\cosh(kz)\cos\phi\right.\nonumber\\
& &\mbox{}\times
\left.\frac{e^{-kd}
(\mu_r^{-1}
\sqrt{k^2+\lambda_L^{-2}}-k)}{k\cosh(kd)+\mu_r^{-1}
\sqrt{k^2+\lambda_L^{-2}}\sinh(kd)}\right)
\end{eqnarray}
for $\Phi_M(\veca{r})$ and 
\begin{eqnarray}
\label{eq21}
g_+(\rho,\phi,z)&=&-\frac{m}{4\pi\mu_r\lambda_L^2}
\int^{\infty}_0\,dk\,k\,e^{-\sqrt{k^2+
\lambda_L^{-2}}(z-d)}J_1(k\rho)
\cos\phi\nonumber\\
& &\mbox{}\times\frac{1}{k\cosh(kd)+\mu_r^{-1}
\sqrt{k^2+\lambda_L^{-2}}\sinh(kd)}
\end{eqnarray}
for $g_+(\veca{r})$ with $g_-(\rho,\phi,z)=-g_+(\rho,-\phi,-z)$
and where $J_1(x)$ is the Bessel function of order one.
These integrals can only be calculated explicitly
in the particular case $d\rightarrow 0$, $\mu_r=1$. We obtain
\begin{eqnarray}
\label{eq22}
\Phi_M(\rho,\phi,0)&=&
\frac{m}{4\pi}\left(\frac{1}{\rho\lambda_L}+
\frac{e^{-\rho/\lambda_L}}{\rho^2}\right)\cos\phi\\
g_+(\rho,\phi,0)&=&-\frac{m\cos\phi}{4\pi\lambda_L^2\rho}\, .
\label{eq23}
\end{eqnarray}
The magnetic potential $\Phi_M(\rho,\phi,0)$ corresponds to
the magnetic potential of a dipole which 
produces a 3d field at short distances 
and that at distances $\gg\lambda_L$ produces a 2d field, i.e.
the field produced by a dipole in a two dimensional space. 
This behaviour can be traced to the Meissner effect, which
confines the flux lines to the spatial gap and to a region of size
$\lambda_L$ in each of the superconductors. 
Although the $d=0$ case 
is somewhat unphysical (the superconductor would have to withstand
an infinite field), we shall see 
that as long as $d\ll\lambda_L$, this type
of behaviour is essentially unchanged.

Now, in order to generalise this solution to the case
of an arbitrary distribution of magnetisation,
we represent the magnetisation per unit of area in the form
\begin{equation}
\veca{m}(\vecg{\rho})=
\sum_{i}\veca{m}_{i}\,\delta^2(\vecg{\rho}-\vecg{\rho}_i)
\label{eq24}
\end{equation}
where $\veca{m}_i$ is a dipole 
situated at $\vecg{\rho}_i=(x_i,y_i)$ and
$\vecg{\rho}=(x,y)$. In real systems, 
$\vecg{\rho}_i$ will correspond
to the sites of a two dimensional 
lattice where the dipoles are situated. 
The solution corresponding to this distribution
of magnetisation is given by the linear superposition of the
solutions corresponding to each  $\veca{m}_i$, i.e.
\begin{eqnarray}
\label{eq25}
\Phi_M(\rho,\phi,z)&=&\sum_{i}\frac{\veca{m}_i\cdot(\vecg{\rho}-
\vecg{\rho}_i)}{4\pi(\mid\vecg{\rho}-
\vecg{\rho}_i\mid^2+z^2)^{3/2}}\\
& &\mbox{}+\sum_{i}\frac{\veca{m}_i\cdot(\vecg{\rho}-
\vecg{\rho}_i)}{4\pi\mid\vecg{\rho}-
\vecg{\rho}_i\mid}\int^{\infty}_0\,dk\,k\,
J_1(k\mid\vecg{\rho}-\vecg{\rho}_i\mid)\,\cosh(kz)\nonumber\\
& &\mbox{}\times\frac{e^{-kd}
(\mu_r^{-1}\sqrt{k^2+\lambda_L^{-2}}-k)}{k\cosh(kd)+\mu_r^{-1}
\sqrt{k^2+\lambda_L^{-2}}\sinh(kd)}\nonumber
\end{eqnarray}
for $\Phi_M(\veca{r})$ and 
\begin{eqnarray}
\label{eq26}
g_+(\rho,\phi,z)&=&-\sum_i\frac{\veca{m}_i\cdot(\vecg{\rho}-
\vecg{\rho}_i)}{
4\pi\mu_r\lambda_L^2\mid\vecg{\rho}-\vecg{\rho}_i\mid}
\int^{\infty}_0\,dk\,k\,J_1(k\mid\vecg{\rho}-\vecg{\rho}_i\mid)\,
e^{-\sqrt{k^2+\lambda_L^{-2}}(z-d)}
\nonumber\\
& &\mbox{}\times\frac{1}{k\cosh(kd)+\mu_r^{-1}
\sqrt{k^2+\lambda_L^{-2}}\sinh(kd)}
\end{eqnarray}
for $g_+(\veca{r})$. The function $g_-(\veca{r})$ 
is constructed from the single dipole
solution in an analogous manner. 
Note that one cannot use the equations
(\ref{eq12}) and (\ref{eq13}) because we no longer
have the $\pi$ rotation symmetry 
around the $x$ axis \cite{note2}. Using the magnetic
potential $\Phi_M(\veca{r})$, we 
can now compute the dipolar energy of
the system.
\section{The magnetic energy of the system}
\label{secD}
The dipolar energy of the system can be obtained by substituting
the dipole distribution by an equivalent distribution of loops
of current, i.e. one which will produce 
the same field distribution. 
The easiest way to compute the energy necessary for the 
formation of such a current distribution is to compute it 
with the currents of the individual
loops which form the distribution kept constant. 
The dipolar energy 
of the magnetisation distribution is equal to the energy necessary 
to form the current distribution when the 
fluxes in each loop are kept
constant, which is minus the energy computed with constant currents
and is given by \cite{Schwartz87,note3}    
\begin{equation}
E_m=-\frac{1}{2}\mu_0\int\,d^2\rho\,\,
\veca{m}(\vecg{\rho})\cdot\veca{h}
(\rho,\phi,z=0)\, .
\label{eq27}
\end{equation}
Substituting $\veca{h}=-\nabla\Phi_M$, with $\Phi_M(\veca{r})$
given by (\ref{eq25}) and approximating 
the discrete sums on the lattice
by integrals, we obtain
\begin{eqnarray}
\label{eq28}
E\!&=&\!\frac{\mu_0}{8\pi}\int_{BZ}\frac{d^2k}{(2\pi)^2}
\frac{2\pi}{k}(\veca{k}\cdot\veca{m}(\veca{k}))(\veca{k}\cdot
\veca{m}(-\veca{k}))\times\\
& &\frac{(\mu_r^{-1}\sqrt{k^2+\lambda_L^{-2}}+k)e^{kd}+
(\mu_r^{-1}\sqrt{k^2+\lambda_L^{-2}}-k)e^{-kd}}{(
\mu_r^{-1}\sqrt{k^2+
\lambda_L^{-2}}+k)e^{kd}-
(\mu_r^{-1}\sqrt{k^2+\lambda_L^{-2}}-k)e^{-kd}},\nonumber
\end{eqnarray}
where $\veca{m}(\veca{k})=\sum_i\,
\veca{m}_i\,e^{-i \vecb{k}\cdot\vech{\rho}_i}$
and where the integrals over $k$ 
are over the first Brillouin zone of the
reciprocal lattice. The expression (\ref{eq28}) 
was obtained in the limit
in which one can disregard the 
lattice structure of the dipole distribution.
If such a structure has to be taken into account, 
then one has to use Ewald summation methods 
\cite{Ewald21} to handle 
the discrete sums in equation (\ref{eq25}).
 
The expression (\ref{eq28}) has two important limits. 
The first is when $\mu_r=1$ and $\lambda_L\rightarrow\infty$,
or when $d\rightarrow\infty$. We obtain
\begin{equation}
E=\frac{\mu_0}{8\pi}\int_{BZ}\frac{d^2k}{(2\pi)^2}\,
\frac{2\pi}{k}(\veca{k}\cdot\veca{m}(\veca{k}))(
\veca{k}\cdot\veca{m}(-\veca{k}))
\label{eq29}
\end{equation}
which is the familiar result 
for the dipolar energy of a thin film. 
The second limit is when the 
largest contribution to the energy comes
from modes $\veca{m}(\veca{k})$ for which 
$k\leq L^{-1}$ where $L$ is a length such that
$L\gg\lambda_L\gg d$ (we take $\lambda_L\gg d$). 
In this case, we obtain
\begin{eqnarray}
E&\approx &\frac{\mu_0}{4\mu_r\lambda_L}\int_{k\leq L^{-1}}
\frac{d^2k}{(2\pi)^2}\,
\frac{(\veca{k}\cdot\veca{m}(\veca{k}))(
\veca{k}\cdot\veca{m}(-\veca{k}))}{k^2}
\left(\,1+\frac{1}{2}(k\lambda_L)^2+O(k^4)\,\right)\,
\label{eq30}
\end{eqnarray} 
which shows that for $k\ll\lambda_L^{-1}$ the energy of the system
has the same form as the energy of a 
system of dipoles in a two dimensional
space. Also, comparing equation
(\ref{eq28}) (with $\mu_r=1$) 
with (\ref{eq29}), one can 
conclude that the dipolar interaction has 
effectively been enhanced with respect 
to the simple film situation,
since the fraction in (\ref{eq28}) 
is always larger than 1 when $\mu_r=1$. 
This result can be easily understood from the fact
that the energy given in
(\ref{eq28}) also includes the kinetic
energy of the supercurrent \cite{note4}. 

\begin{figure}[htbp]
\centerline{\epsfxsize 0.75\columnwidth \epsfbox{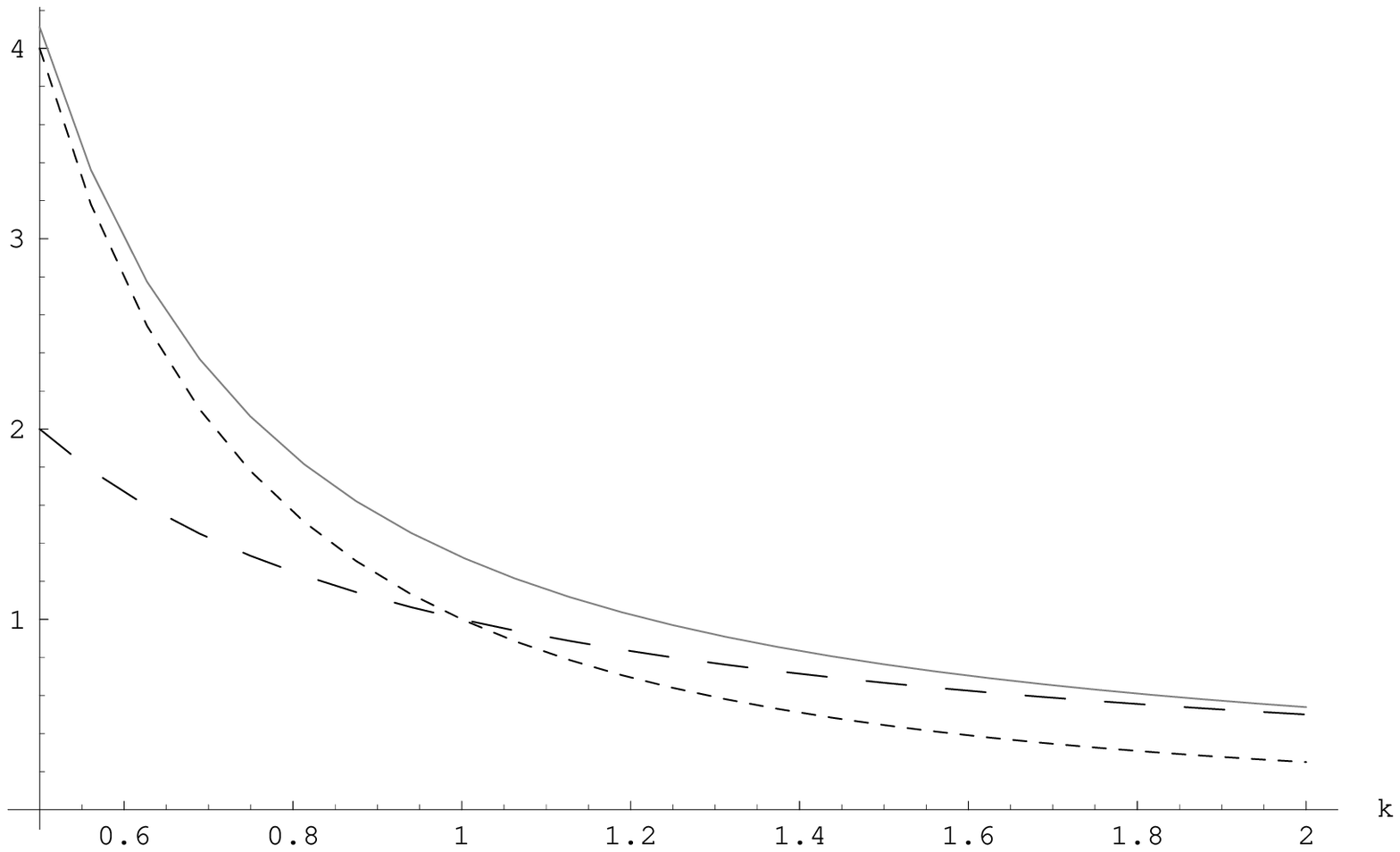}}
\vspace{3mm}
\caption{\label{F3}}
\end{figure}

The modified dipolar kernel $F(k)=\frac{1}{k}\,
\frac{(\sqrt{k^2+\lambda_L^{-2}}+k)e^{kd}+
(\sqrt{k^2+\lambda_L^{-2}}-k)e^{-kd}}{(
\sqrt{k^2+
\lambda_L^{-2}}+k)e^{kd}-
(\sqrt{k^2+\lambda_L^{-2}}-k)e^{-kd}}$
given in (\ref{eq28}) with $\mu_r=1$,
is plotted against the kernels $1/k$ and $1/k^2$,
which appear respectively, in (\ref{eq29}) and
(\ref{eq30}), in figure \ref{F3}.

The question now arises if one can 
indeed detect such change of behaviour
in the dipolar interaction 
in artificial superlattices of superconductors
and ferromagnetic materials or even in naturally layered
systems like RuSr$_2$\-Gd\-Cu$_2$\-O$_{8-\delta}$ or 
RuSr$_2$Gd$_{1+x}$\-Ce$_{1-x}\-$Cu$_2$O$_{10}$. 
The detection of such a change of behaviour                  
relies on the possibility of 
finding an experimental system 
which obeys a series of constraints. 

Firstly, the London appro\-xi\-ma\-tion, 
which was used and which
postulates a local relation between the current 
and the  magnetic induction, is only valid sufficiently 
close to the critical temperature at which 
the superconducting phase transition occurs.
But we believe that the qualitative features of this solution 
should be valid even when the London
approximation is not (i.e. at low 
temperatures compared to the critical
temperature). Physically, one should expect this type of behaviour
as long as the superconductor displays the Meissner effect.

Secondly, we have considered a ferromagnetic layer with
zero thickness. The solution obtained
can only be valid if one can neglect 
the proximity effect in the ferromagnetic layer, 
i.e. the induction of superconductivity in the
ferromagnetic material by the superconductor \cite{Genn69}.
Otherwise, in a layer with finite thickness,
an effective exchange interaction 
will be induced between the magnetic moments
and this interaction also has to be taken into account. 
However, this effect is negligible in insulators. 
Therefore, an appropriate material to choose
for the ferromagnetic layer would be a ferromagnetic insulator
\cite{PRainer}.
We have also neglected the supression of the 
superconducting order parameter 
at the boundary between the two media, 
which always occurs in the presence of a 
film of finite thickness.    
In the case of a film made of a ferromagnetic insulator, 
the de Gennes boundary conditions tell us
that the order parameter is effectively quenched to zero at the
boundary \cite{Genn64}. Since the London penetration depth depends
on this parameter, 
this will mean that the effective London
penetration depth will be greater 
than the penetration depth measured
in the absence of the ferromagnetic film. Nevertheless, the London
approximation remains valid (provided 
that we are working in a weak 
field situation, see below). The opposite limit, 
in which the coupling between the superconducting and
the ferromagnetic layers is primarily due to 
the proximity effect has
been considered, using a microscopic approach
by Radovi\'c et al. \cite{Radovic88} and by Schinz
and Schwabl \cite{Schinz92} using a phenomenological description.
In both cases, the problem treated 
reduces to the one of decoupled thin
superconducting layers embedded in a ferromagnet. Experiments done
in superlattices of Fe/V \cite{Wong86} and Nb/Gd \cite{Strunk94} 
have confirmed these results. 

Thirdly, one cannot 
have any Josephson currents flowing through the 
insulating junction,  otherwise the boundary condition 
$\jmath_z=0$ at $z=\pm d$ is not valid. This implies that the 
phase difference between the superconducting order parameters 
of the two superconductors is zero. 
This is the case for most superconducting systems in equilibrium 
but in certain junctions containing ferromagnetic materials a 
non-zero difference between the phases of the superconducting 
order parameter in each side of the junction, 
leading to a current in equilibrium, 
has been predicted \cite{Bulae77}.

Finally, we have not considered the thermodynamical
stability of the system we are working with. This implies
that we are working in a weak field situation, 
i.e. the field produced
by the distribution of magnetisation 
is much lower than the critical 
field at which the superconducting system undergoes a transition
to the intermediate state (in the case 
of type I superconductors) or
to the mixed state (in the case of type II superconductors).
One way to achieve this condition is 
to choose either a magnetic material
with a low saturation magnetisation or a superconductor with a 
high critical field (in the case of type 
I superconductor), e.g. V, 
or a high field $H_{c1}$ (in the case of a type II superconductor),
e.g. Nb. 

In summary, despite the restrictions pointed above,
we think that there is room 
for believing that one could manufacture systems composed
of alternating magnetic and superconducting layers 
where the above effect
could be detected using the methods discussed in the 
introduction or others.
\section{Conclusions}
\label{secE}
We have computed the magnetic field distribution of an
arbitrary distribution of planar
magnetisation in a spatial gap between two superconductors.
The purpose of this calculation is to provide 
a simple model for a superlattice of ferromagnetic
materials and superconductors where the dipolar interaction
between the magnetic moments in the ferromagnetic
system is taken into account.

We have also computed the dipolar energy of such a system and we
have shown that for low momenta compared to the London 
inverse penetration depth, the system has a dipolar energy which 
resembles the energy of a magnetisation 
distribution in 2 dimensions
and hence it leads to an enhancement of the dipolar energy compared
to that of a simple ferromagnetic film.

As to possible future directions of 
research one can point out the 
possibility to consider a treatment 
of the same problem in the line of what
was done in reference \cite{Schinz92}, 
i.e. consider the Landau-Ginzburg
theory of the layered system 
including the free energy of the superconducting
system and of the ferromagnetic layer 
and the interaction between the two 
via the electromagnetic interaction 
and the proximity effect, with the 
appropriate boundary conditions. 
However, in this case one has to take into
account the non-linearity of the problem and the three dimensional 
character of a solution involving a 
non-uniform distribution of magnetisation
in the plane. 

A slightly different model which can also be treated 
exactly with the methods developed in this paper is the
one of a thin ferromagnetic layer between two bulk 
type II superconductors above $H_{c1}$, i.e.
with flux penetration in the
form of an Abrikosov vortex lattice oriented along the
$z$ axis. In this case, the boundary conditions
(\ref{eq17}-\ref{eq19}) would be unchanged, but the
equation determining $g(\veca{r})$, equation (\ref{eq9}), would
be replaced by an inhomogeneous equation where the
role of sources is played by the vortices. Also, the
$\pi$ rotational invariance of the system along the
$x$ axis is no longer present, but one can still
draw useful conclusions from the reflexion properties
of the system in the $xy$ plane (see \cite{note2}).
What makes the model much more difficult 
to solve is the fact that one
cannot exclude the presence of Neumann functions 
in the Fourier-Bessel integrals which determine
$g(\veca{r})$ and $\Phi_M(\veca{r})$, and also that in this case 
one cannot use the equation
(\ref{eq27}) as the expression for the total
magnetic energy of the system (compare with \cite{note3}).
One should nevertheless mention that a  method for computing the magnetic
energy in the presence of vortices has been considered
by Erdin et al. \cite{Erd00} for the case of a thin magnetic
film interacting with a thin superconducting film.
\\ \\
{\bf Acknowledgements:}
We acknowledge many helpful dis\-cus\-sions 
with M. Kuli\'{c}, D. Rainer, H. Braun,
H. Kinder, G. Eilenberger, M. A. Santos, C. Bracher, M. Riza
and M. Kleber. 
J.E.S. acknowledges the support of the 
European Co\-mmis\-sion,
Contract No. ERB\--FMBI\--CT 97\--2816
and from the Deuts\-che Fors\-chungs\-gemein\-schaft
Schwerpunktprogramm `Strukturgradienten in
Kristallen', contract no. Schw 348/12-1 (from 01/03/00). 
E. F. acknowledges the 
support of the Deuts\-che Forschungsgemeinschaft 
through an Heisenberg fellowship, contract no. FR850/3.
E. F. also acknowledges the hospitality of the 
Institut f\"ur Theoretische Physik, LMU M\"unchen, 
where part of this work was done.
F. S. acknowledges the support of the 
Deuts\-che Fors\-chungs\-gemein\-schaft Einzelprojekt
Schw. 348/10-1 and of the 
BMBF Verbundprojekt 03-SC5-TUM0.

\newpage
\noindent{{\bf  Figure Captions}\\
{\bf Figure 1.}
Thin ferromagnetic film between
bulk superconductors (schematic).\\
{\bf Figure 2.}
In-plane distribution of magnetisation
between bulk superconductors (schematic).}\\
{\bf Figure 3.}
The modified dipolar kernel $F(k)$ (the
continous grey plot) is plotted against the kernels
$1/k$ (the long-dashed plot) and $1/k^2$ (the short-dashed
plot). We have taken $\mu_r=1$ and $d=0.1$, $\lambda_L=1$ in $F(k)$
(in arbitrary units). It is seen that the function $F(k)$
interpolates between $1/k^2$ and $1/k$, the crossover
length being the London penetration depth $\lambda_L=1$.
\newpage
\newpage
\newpage

\begin{thebibliography}{}
\bibitem{note1} Note that in this paper we have considered,
for simplicity, a system of planar magnetic moments (an XY model,
in the language of field theory). In some ferromagnetic systems
this constraint is obeyed but in general one would have to consider
a magnetic moment with an arbitrary orientation with respect to the
$xy$ plane. Due to the linearity of the London equations, such a
problem can be splited into the problem which we have solved and
the problem of a magnetic moment oriented along the $z$ axis between two 
superconductors.
\bibitem{note2} One can, instead of 
using the $\pi$ rotation invariance
around the $x$ axis, use the reflexion 
invariance of the problem in
the $xy$ plane, which is also valid 
in the multiple dipole case. But one
has to remember that $\veca{h}$, $\veca{b}$ 
and $\veca{m}(\vecg{\rho})$
are axial vectors, which makes the analysis more complicated.
\bibitem{note3} It can also be shown 
from the general expression for
the total energy of a system of 
superconductors treated in the
London approximation \cite{London61} 
and of the ferromagnetic layer
\cite{Landau60}, 
using equations (\ref{eq1}) to (\ref{eq6}) 
and the boundary conditions
for $\veca{h}$ and $\veca{b}$, that the only term which survives is 
the one given by equation (\ref{eq27}).
\bibitem{note4} It can be easily shown that the difference
between equation (\ref{eq28}) at finite $\lambda_L$ and
at infinite $\lambda_L$ is always positive, which shows
that the magnetic energy of the system is always larger
in the superconducting state.
\end{thebibliography}
\end{document}